\documentclass[aps,twocolumn,showpacs,amsmath,amssymb,superscriptaddress,floatfix,final,10pt,prb]{revtex4-1}

\usepackage{graphicx}
\usepackage{color}
\usepackage{dcolumn} 
\usepackage{bm}      
\usepackage{nicefrac}
\usepackage{mathrsfs}

\newcommand{\BACSO}{Ba\-Cu$_2$\-Si$_2$\-O$_7$}
\newcommand{\BACGO}{Ba\-Cu$_2$\-Ge$_2$\-O$_7$}
\newcommand{\BASGE}{Ba\-Cu$_2$\-Si\-Ge\-O$_7$}

\begin{document}
\title{Impact of strong disorder on the static magnetic properties of the spin-chain compound BaCu$_2$SiGeO$_7$}

\author{T.\ Shiroka}
\affiliation{Laboratorium f\"ur Festk\"orperphysik, ETH H\"onggerberg, CH-8093 Z\"urich, Switzerland}
\affiliation{Paul Scherrer Institut, CH-5232 Villigen PSI, Switzerland}

\author{F.\ Casola}
\email{fr.casola@gmail.com}
\affiliation{Laboratorium f\"ur Festk\"orperphysik, ETH H\"onggerberg, CH-8093 Z\"urich, Switzerland}
\affiliation{Paul Scherrer Institut, CH-5232 Villigen PSI, Switzerland}

\author{W.\ Lorenz}
\affiliation{Neutron Scattering and Magnetism, Laboratorium f\"ur Festk\"orperphysik, ETH H\"onggerberg, CH-8093 Z\"urich, Switzerland}

\author{A.\ Zheludev}
\affiliation{Neutron Scattering and Magnetism, Laboratorium f\"ur Festk\"orperphysik, ETH H\"onggerberg, CH-8093 Z\"urich, Switzerland}

\author{H.-R.\ Ott}
\affiliation{Laboratorium f\"ur Festk\"orperphysik, ETH H\"onggerberg, CH-8093 Z\"urich, Switzerland}
\affiliation{Paul Scherrer Institut, CH-5232 Villigen PSI, Switzerland}

\author{J.\ Mesot}
\affiliation{Laboratorium f\"ur Festk\"orperphysik, ETH H\"onggerberg, CH-8093 Z\"urich, Switzerland}
\affiliation{Paul Scherrer Institut, CH-5232 Villigen PSI, Switzerland}

\date{\today}

\begin{abstract}
The disordered quasi-1D magnet BaCu$_2$SiGeO$_7$ is considered as 
one of the best physical realizations of the random Heisenberg chain 
model, which features an irregular distribution of the exchange parameters 
and whose ground state is predicted to be the scarcely investigated 
\textit{random-singlet} state (RSS). 
Based on extensive $^{29}$Si NMR and magnetization studies of BaCu$_2$SiGeO$_7$, 
combined with numerical Quantum Monte Carlo simulations, we obtain remarkable 
quantitative agreement with theoretical predictions of the random Heisenberg 
chain model and strong indications for the formation of a random-singlet state 
at low temperatures in this compound.
As a local probe, NMR is a well-adapted technique for studying the magnetism of 
disordered systems. In this case it also reveals an additional local transverse 
staggered field (LTSF), which affects the low-temperature properties of the RSS. 
The proposed model Hamiltonian satisfactorily accounts for the temperature 
dependence of the NMR line shapes.
\end{abstract}

\pacs{75.10.Pq, 76.60.-k, 75.10.Jm, 75.40.Cx}


\maketitle

\section{\label{Intro}Introduction}
Spin-\nicefrac{1}{2} Heisenberg chains adopt a non-magnetic ground state, which qualitatively 
can be seen as a linear superposition of states representing all the possible ways of forming 
singlets in the system.\cite{BetheA} 
With quantum fluctuations suppressing any long-range order, the translational symmetry is 
preserved. At large length scales (low energies) \textit{any} amount of disorder in the 
exchange parameters is predicted to dominate over the quantum or thermal 
fluctuations\cite{Dasgupta79,Fisher94,Igloi05} and the resulting system is known as 
``random Heisenberg chain'' (RHC). In practice, the disorder leads to an inhomogeneous 
ground state by associating to every random configuration of exchange paths a unique 
way of forming singlets, regardless of the distance between the involved spins and of 
their interactions. 
This new type of ground state, specific to random Heisenberg chains, is called 
\textit{random-singlet state} (RSS).\cite{Fisher94}

It is widely accepted\cite{Igloi05} that the interest in RHCs originally arose from a 
novel approach to deal with random exchanges in the isotropic spin-\nicefrac{1}{2} 
Heisenberg model, introduced in 1979 by Dasgupta and Ma.\cite{Dasgupta79} Their 
new physical insight into the effects of disorder was later developed into a formal 
theory by Fisher\cite{Fisher94} and, subsequently, applied to a large variety of problems 
involving magnets with quenched disorder.\cite{Igloi05,Andrade12,Lin03,Laflorencie05}
However, it is only the recent combined availability of materials representing physical 
realizations of disordered quasi-1D quantum magnets\cite{Zheludev13,Ward13,Simutis13} 
and of novel stochastic numerical methods\cite{Bauer2007} that made possible 
the first quantitative studies concerning the impact of disorder on materials whose 
magnetic properties are well described by an antiferromagnetic Heisenberg-chain 
Hamiltonian [see Eq.~\eqref{rhc1}].

\begin{figure}[bth]
\centering
\includegraphics[width=0.42\textwidth]{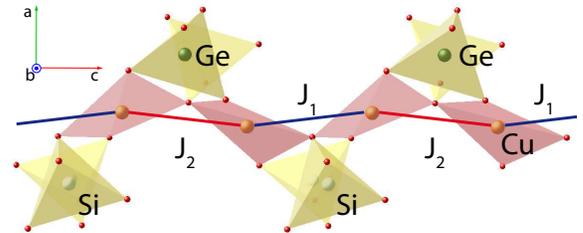}  
\caption{\label{fig:struct_BASGE}(color online) Sketch of the relevant
structural unit in BaCu$_2$SiGeO$_7$ highlighting the different $J_1$ 
and $J_2$ exchange couplings. Spin chains run along the crystallographic 
$c$-axis.} 
\end{figure}

The aim of the present work was to identify a model material featuring the 
properties of an RHC and, through experimental and numerical methods, 
to demonstrate that the chosen RHC Hamiltonian describes its physical behavior. 
Here we present a set of field- and temperature-dependent magnetization 
and ${}^{29}$Si nuclear magnetic resonance (NMR) data of the electronic 
insulator \BASGE.\cite{Yamada01} Its relevant structural unit is shown in 
Fig.~\ref{fig:struct_BASGE} and its physical properties can be discussed in terms 
of the one-dimensional Hamiltonian:
\begin{align}
\mathscr{H} &= \sum_i \left[J_{xy}(i) \left( S^x_iS^x_{i+1}+S^y_iS^y_{i+1} \right) + J_{z}(i) S^z_iS^z_{i+1} \right] \nonumber \\
& -  g \mu_{\mathrm{B}} \sum_{i}H S^z_i +  \mu_{\mathrm{B}} H_{\perp} \sum_{i} (-1)^i S^x_i. \label{rhc1}
\end{align}
Here $J_{xy}(i)=J_{z}(i)=J_i>0$ represent the random exchange couplings along 
the spin-chain sites $i$, $\mu_{\mathrm{B}}$ is the Bohr magneton, $H$ the uniform 
applied magnetic field, $H_{\perp} = cH$ a locally induced transverse staggered 
field,\cite{Casola12} and the spin operators refer to a spin $S=1/2$.

In order to achieve a description of the large length-scale/low-energy physics contained 
in Eq.~\eqref{rhc1} with $H=0$, Dasgupta and Ma developed a \textit{real-space} 
renormalization procedure. They started from the local bonds rather than from the 
large spatial blocks, as is usually done for homogeneous systems.\cite{Kadanoff76}
Starting from the distribution of unrenormalized exchange couplings $P(J_i,J)$, where 
$J$ is the energy cutoff, an effective low-energy theory is constructed by tracking, 
via flow equations, the behavior of $P(J_i,J)$ as the cutoff $J\sim k_{\mathrm{B}}T$ is 
progressively reduced. 
In the RSS theory, the flow of the distribution of unrenormalized bond energies $P(J_i)$ 
depends on the considered energy scale and temperature.\cite{Dasgupta79,Fisher94} 
As a result, the static magnetic properties are dominated by a temperature-dependent 
population of effectively paramagnetic moments $n(T)$. Their characteristic 
energy- ($\omega$) and temperature-dependent average spatial separation 
$\xi(T,\omega=0) \sim n^{-1}(T) = \log^2(J_0/k_{\mathrm{B}}T)$ diverges as $T \rightarrow 0$, 
leading to a quasi long-range ordered ground state.\cite{Fisher94} In the expression 
for $n^{-1}(T)$, $J_0$ is the largest exchange coupling in the chain prior to the 
renormalization.\cite{Shiroka11}
Due to weak residual interchain  interactions, a magnetic order at a N\'eel temperature 
$T_{\mathrm{N}}>0$ K is established also in the random Heisenberg chains, albeit with a 
lower ordering temperature in comparison with the disorder-free case.\cite{Thede12,Yusuf05} 
In the present work we focus our attention on the local and bulk static magnetic properties 
of \BASGE\ for $T>T_{\mathrm{N}}$ and $H \neq 0$.

\section{B\lowercase{a}C\lowercase{u}$_2$S\lowercase{i}G\lowercase{e}O$_7$: Summary of previous research}
\label{sec:PreviousRes}
The experimental work on \BASGE\ began with studies on tunable super-exchange interactions 
in spin-chain systems by Yamada \textit{et al}. [\onlinecite{Yamada01}].
Across the series from \BACSO\ to \BACGO\ the BaCu$_2$(Si$_{1-x}$Ge$_x$)$_2$O$_7$ 
compounds ($0<x<1$) crystallize with a \textit{Pnma} space group. For $x=0.5$, corresponding 
to maximum disorder in the exchange couplings, the values of the lattice parameters are 
$a=6.917(7)$ \AA, $b=13.28$ \AA, and $c=6.944(7)$ \AA,\cite{Zheludev07} in between the 
parameters of the two end members of the series. The spin chains run along the 
crystallographic $c$ axis, as shown in Fig. \ref{fig:struct_BASGE}.

Early magnetization measurements of polycrystalline samples revealed the often observed broad 
maximum of $\chi(T)$, also known as the Bonner-Fisher peak.\cite{Bonner64} This maximum 
shifts linearly with $x$ towards higher temperatures as the Ge concentration is enhanced. 
From this feature it was initially deduced that the high-temperature properties of \BASGE\ 
effectively reflect those of a standard spin-\nicefrac{1}{2} Heisenberg chain with 
$J_{\mathrm{eff}} = (J_{\mathrm{Si}} + J_{\mathrm{Ge}})/2 \approx 37$ meV, 
where $J_{\mathrm{Si}} = 24.1$ meV and $J_{\mathrm{Ge}} = 50$ meV represent the intrachain 
exchange interactions for no and for complete Ge substitution, respectively. The enhancement 
of the AFM exchange, resulting from a higher Ge content, was explained in terms of a 
change of the Cu-O-Cu  bonding angle, from $\phi$ = 124$^{\circ}$ in \BACSO\ to 135$^{\circ}$ 
in the Ge parent compound.\cite{Bertaina06} 

Earlier bulk measurements already revealed two distinct features:\cite{Yamada01} (\textit{i}) a 
low-temperature divergence of the spin susceptibility in \BASGE, interpreted as a simple Curie-type 
behavior due to uncompensated magnetic moments,\cite{Masuda04} and (\textit{ii}) a low-temperature 
magnetic order, expected even in the case of random Heisenberg chains (RHCs),\cite{Yusuf05} 
with an onset at $T_{\mathrm{N}} = 0.7$ K in \BASGE.\cite{Zheludev07} 

In zero magnetic field, \BACSO\ is known to order antiferromagnetically at 
$T_{\mathrm{N}} = 9.2$ K.\cite{Kenz01}
The more than tenfold reduction of $T_{\mathrm{N}}$ in \BASGE, may be related to the fact 
that the introduction of Ge modifies not only the intrachain, but also the interchain coupling 
which, along the $a$-axis, changes from FM- to AFM-type.\cite{Yamada01} 
The latter implies that for \BASGE\ the interchain coupling alternates in sign and, therefore, 
it averages out to a mean-field value.\cite{Joshi03}

The interest in \BASGE\ grew significantly as soon as results of magnetic-susceptibility
measurements on single crystals appeared in the literature.\cite{Masuda04} 
They suggested a logarithmic dependence of the low-temperature spin susceptibility,\cite{Masuda04} 
consistent with predictions of the RHC model.\cite{Dasgupta79} 
Subsequent inelastic neutron-scattering (INS) studies focused on the measurement of the 
energy dependence of the correlation length $\xi(T=0,\omega)$, which at zero-temperature 
is inversely proportional to the free-spin concentration 
$\xi(\hbar \omega) \sim n^{-1}(\hbar \omega) = \log^2(J_0/ \hbar \omega)$, with 
$\hbar \omega$ the energy of the probing neutrons. 
In INS, $\xi$ is extracted from the inverse width of the energy-integrated scattering intensity 
$S(\mathbf{q})$ which, in case of a disordered phase, is typically of Lorentzian shape.\cite{Birgeneau99} 
Following some earlier misinterpretation of the data due to problems with background 
subtraction,\cite{Masuda04,MasudaErr} the INS results were found to be quantitatively consistent 
with a Luttinger-liquid behavior reflecting a disorder-free spin chain with a single effective 
coupling $J_{\mathrm{eff}} \approx 37$ meV.\cite{Zheludev07}
It was concluded that the expected RHC-related physics must manifest itself at energies much 
lower than those accessible by neutron experiments. In a previous NMR study\cite{Shiroka11} 
we already pointed out how the spin dynamics of \BASGE, probed at the characteristic Larmor 
energy $\hbar \omega_{\mathrm{L}} \sim 1$ $\mu$eV (with $\omega_{\mathrm{L}}/2 \pi$ the 
NMR resonance frequency), turns out to be notably different with respect to that of the regular 
Heisenberg chain \BACSO. 
On the other hand, considering existing data on \BACSO\ and \BACGO, a low-temperature 
divergence of the magnetic susceptibility also in \BASGE\ is not really surprising.\cite{Casola12} 
In addition, since INS failed to detect any sort of disorder-induced spin dynamics in this compound,  
the existence of a random-singlet state (RSS --- the ground state of the isotropic RHC Hamiltonian) was 
considered  questionable. In the present work we show how the effect of disorder on the 
\textit{static} magnetic properties can be quantitatively modeled and compared with experiments, 
thereby supporting the RHC scenario.

\section{Magnetic susceptibility of B\lowercase{a}C\lowercase{u}$_2$S\lowercase{i}G\lowercase{e}O$_7$}  
\label{sec:magnQMC}
\subsection{Experimental results}
\begin{figure}[bth]
\centering
\includegraphics[width=0.45\textwidth]{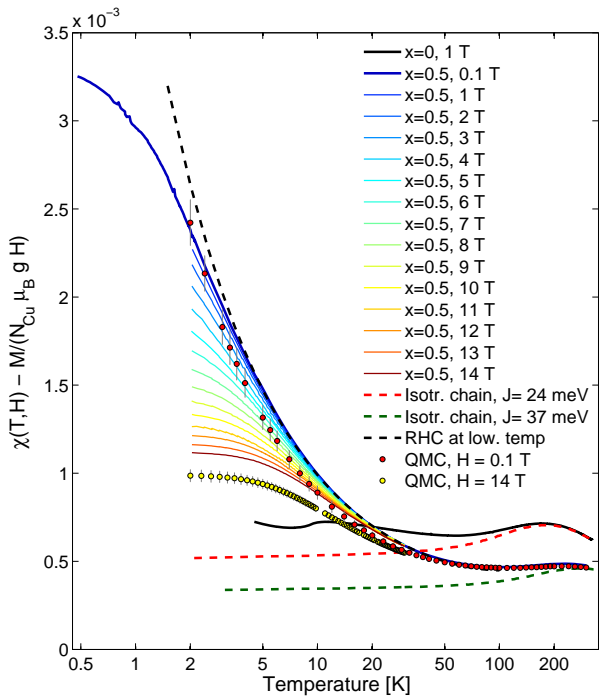}  
\caption{\label{fig:RHC_chi_meas}(color online) Field and temperature dependences of the magnetic 
susceptibility $\chi(T,H)$ per Cu ion in \BASGE\ (and in \BACSO, at 1 T) with $\mathbf{H} \parallel b$ 
in normalized units, assuming $g = 2$.\cite{Casola12} The fit of $\chi(T,0.1\,\mathrm{T})$ with an 
RHC model was adopted from Ref.~\onlinecite{Fisher94}, while the $\chi(T)$ curves for chains with 
a single isotropic exchange value were calculated as in Ref.~\onlinecite{Johnston00}. QMC 
simulations of the $\chi(T,H)$ susceptibility at $\mu_0 H=0.1$ and 14 T for a simple $J_1$--$J_2$ 
RHC model is also shown (see text for details).}
\end{figure}
As a first step, we extended the measurements of the magnetic susceptibility $\chi(T,H)$ 
to wider temperature and field ranges than those covered in previous studies.
Single crystals of \BASGE\ were grown by the floating-zone technique.
One of them was cut and the smaller 6.5-mg part was used in magnetometry measurements; 
the remaining 20-mg piece was used for the NMR investigations.

The magnetization of the chosen sample was measured by applying a magnetic field $\mathbf{H} \parallel b$. 
Magnetic susceptibility data $\chi = M/H$ are displayed as bold solid lines in Fig.~\ref{fig:RHC_chi_meas}.
For $\mu_0 H = 0.1$ T and at temperatures $T \geq 1.8$ K, the measurements were 
performed using a standard dc SQUID magnetometer, whereas to cover temperatures 
down to 0.5 K we employed a SQUID device with a $^3$He insert.
For all the other fields and for temperatures $T \geq 1.8$ K we used a vibrating-sample 
magnetometer (VSM). 
Since in the latter case data are collected ``on the fly'', i.e., while the temperature 
is being swept at 0.25 K/min, this causes a small (known) temperature difference
between sample and thermometer position. In addition, a certain dispersion in the 
magnetization values was reduced by applying a moving-average filter with a 200-mK width 
to the raw experimental data. Magnetic hysteresis, defined as the difference between 
the zero-field-cooled and field-cooled magnetization data, was observed for $T < 800$ mK. 
The hysteretic behavior is probably related to ferromagnetic domains with a rather small 
net magnetic moment present in the ordered low-temperature phase. Since only the 
zero-field-cooled scans are relevant for our discussion, the field-cooled data are not shown.

\subsection{Data analysis}
As shown in Fig. \ref{fig:RHC_chi_meas}, for temperatures above 30 K the $\chi(T,H)$ curves 
are field-independent and, for $T>200$ K, they approach the theoretical prediction for 
an \textit{isotropic} Heisenberg chain \cite{Johnston00} with an exchange parameter 
$J_{\mathrm{eff}} = (J_{\mathrm{Si}} + J_{\mathrm{Ge}})/2 \approx 37$ meV (see the green 
dashed line in Fig. \ref{fig:RHC_chi_meas}).
For comparison, the red dashed line indicates the calculated $\chi(T)$ for an isotropic chain 
with $J_{\mathrm{Si}} \approx 24$ meV, corresponding to the $x=0$ case.

We first discuss two major differences between the magnetic susceptibilities of \BACSO\ 
($x=0$) and \BASGE\ ($x=0.5$):
(\textit{i}) Our previous successful attempt to interpret the low-temperature increase/divergence 
of $\chi(T)$ in \BACSO\ and \BACGO\ as simply being caused by a local transverse 
staggered field (LTSF,\cite{Casola12} see Sec.~\ref{ssec:nmr_analysis}) is inadequate 
for treating the more complex case of \BASGE.
First of all, the magnetic susceptibility $\chi(T < 10\,\mathrm{K},H)$ is strongly \textit{reduced} 
even by moderate fields. As a result, the magnetization deviates from the linear response 
with respect to $H$ (see Fig.~\ref{fig:M_H_qmc} and Ref.~\onlinecite{Johnston00}), 
expected either for an isotropic chain in the LL regime, or for the sine-Gordon (SG) 
model in the high-temperature limit, the latter describing the physics when an LTSF 
is present.\cite{Casola12} 

The SG model, even at intermediate temperatures, is inadequate to describe the physics 
of \BASGE. Indeed, in this case $\chi(T)$ is known to exhibit a symmetric peak centered at a temperature 
$T_{m}$, roughly corresponding to half of the field-induced gap $\Delta_{\mathrm{SG}}$.\cite{Affleck99}
To exemplify this fact, we consider the Hamiltonian in Eq.~\eqref{rhc1} with $J_i = J$ and 
$H \neq 0$. In the reduced units $h^*_{i,\perp} = \mu_{\mathrm{B}} H_{i,\perp}/J_{\mathrm{eff}}$, 
with $H_{i,\perp}=(-1)^icH$, this model features a gap $\Delta_{\mathrm{SG}}$ in the 
spin-excitation spectrum which, for $h^*_{\perp} \ll 1$, scales as:\cite{Shibata01}
\begin{figure}[bh]
\centering
\includegraphics[width=0.48\textwidth]{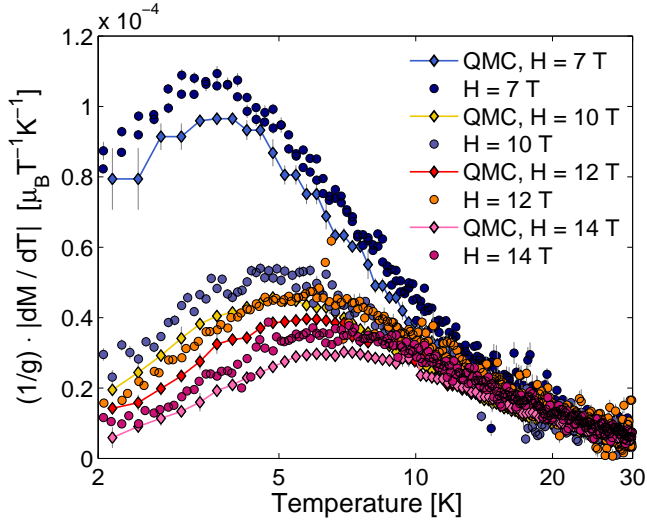}  
\caption{\label{fig:Phase_bound_cuts}(color online) First derivative of the sample magnetization 
vs.\ temperature, $|\partial{M}(T,H)/\partial{T}|$, in \BASGE\ for $H \parallel b$. The graphic 
compares several cross sections of the color map shown in Fig.~\ref{fig:Phase_bound}. The 
assigned errors reflect the line-slope uncertainty for fits within the binning range (see text for 
details). The shown QMC simulation results were obtained \textit{without} using any free parameters.}
\end{figure}
\begin{equation}
\frac{\Delta_{\mathrm{SG}}}{J} = 1.78 \cdot \left( h^*_{\perp} \right)^{\frac{2}{3}} \cdot \left( -\log  h^*_{\perp} \right)^{\nicefrac{1}{6}}. 
\label{gapLTSF}
\end{equation}
Equation~\eqref{gapLTSF} implies that for $c = 0.092$ (the value estimated for \BACSO \cite{Casola12} 
if $\mathbf{H} \parallel b$), the LTSF yields $\Delta_{\mathrm{SG}} \simeq 1.42$ meV at $H=14$ T, 
corresponding to $T_{m} \simeq 8.2$ K. On the other hand, as can be seen in Fig.~\ref{fig:RHC_chi_meas}, 
the susceptibility recorded at 14 T tends to saturate at low temperatures, rather than display a 
peak at $T_{m}$.

\begin{figure}[ht!]
\centering
\includegraphics[width=0.48\textwidth]{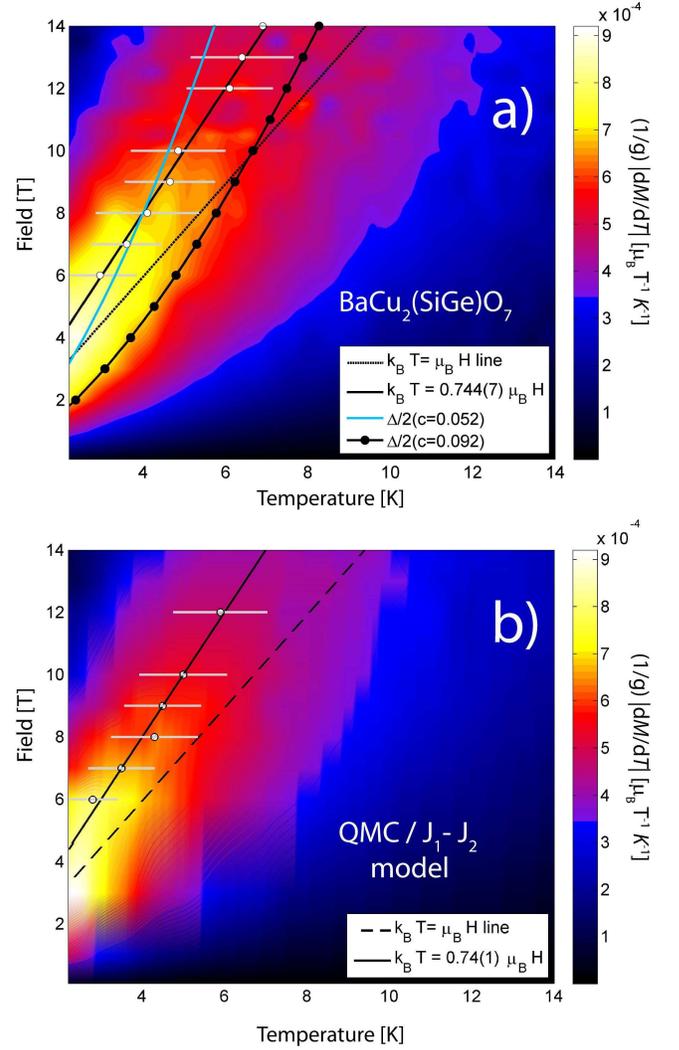}  
\caption{\label{fig:Phase_bound}(color online) (a) First derivative of the sample magnetization 
vs.\ temperature, $|\partial{M}(T,H)/\partial{T}|$, in \BASGE\ for $H \parallel b$ (data scaled 
using $g=2$).\cite{Casola12} The open circles represent cross-over temperatures. 
For details on data treatment and for a description of the plotted lines see text. 
(b) QMC simulation of the $J_1$--$J_2$ RHC model for a system of 6000 spins, 
using 20 realizations of disorder for each $(T,H)$-point. The color map was 
generated using the same data treatment as in panel (a).}
\end{figure}

(\textit{ii}) The inadequacy of the SG model in interpreting the $\chi(T,H)$ data of \BASGE\ 
is particularly evident from the $(T,H)$-dependence of $T_i$, the inflection point which 
marks the start of saturation (corresponding to the maximum of $|\partial{M}/\partial{T}|$ 
in a fixed field $H$ --- see Fig.~\ref{fig:Phase_bound_cuts}). 
To overcome noise problems in the calculation of $|\partial{M}/\partial{T}|$ 
from raw magnetization data, the experimental points were linearly interpolated 
using a 200-mK binning and the resulting line slope was taken as the derivative 
with respect to temperature. The results of this procedure are shown as colored 
circles in Fig.~\ref{fig:Phase_bound_cuts} for four selected fields. 
A $(T,H)$ color map of the $|\partial{M}/\partial{T}|$ values is shown in Fig.~\ref{fig:Phase_bound}a. 
Here the inflection points (assigned with an error corresponding to the peak width at 95\% 
of the peak height in Fig.~\ref{fig:Phase_bound_cuts}) are shown as open circles. 
The resulting $T_i(H)$ values follow a linear field dependence, such that 
$k_{\mathrm{B}} T_i = 0.744(7) \mu_{\mathrm{B}} H$.  
Besides requiring a nonuniversal dimensionless prefactor,\cite{Westerberg95} 
the field dependence of $T_i$ does not follow the $H^{\nicefrac{2}{3}}$ power law 
that the small $h^*_{i,\perp}$ value would imply for $\Delta(H)/2$.\cite{Affleck99} 
The calculated dependences of $\Delta/2$ for $H_{\perp}=(-1)^icH$ with $c=0.092$ 
or 0.052 are both also shown in Fig.~\ref{fig:Phase_bound}a. 
The first $c$ value represents the LTSF proportionality constant in \BACSO.\cite{Casola12} 
Because of the similar local environments, it seems reasonable to use the same 
value also for \BASGE. The reason for choosing $c=0.052$ instead, will be 
clarified in Sec.~\ref{sec:nmr1}.

In view of these two inadequacies, we modified our previous approach to the experimental 
data analysis\cite{Casola12} and instead chose to compare the experimental results with 
the predictions of the RHC model Hamiltonian in Eq.~\eqref{rhc1}.\cite{Fisher94}
A first comparison is carried out by fixing $c=0$ and by assuming the $J_i$ values to be 
uniformly distributed between the two limits, $J_{\mathrm{Si}}$ and $J_{\mathrm{Ge}}$.

In case of nonzero temperatures and magnetic fields, the renormalization flow of 
the decimation procedure is interrupted either at the thermal energy $k_{\mathrm{B}} T$, 
or at the magnetic energy $g \mu_{\mathrm{B}} \langle S \rangle H$.\cite{Westerberg95} 
In the first case, the resulting susceptibility is:\cite{Fisher94}
\begin{equation}
\chi(T)_{\mathrm{RHC}} \sim  T^{-1} \log^{-2}(J_0/k_{\mathrm{B}}T).
\end{equation}
In the second case, the system is driven away from the zero-field fixed point into a 
regime with field-aligned undecimated spins, a saturated magnetization and zero 
entropy.\cite{Westerberg95} 
In spin-$\nicefrac{1}{2}$ AFM-RHCs the decimation procedure involving the total set of 
spins leaves the renormalized magnetization magnitude at $\langle S \rangle = 1/2$, 
unlike, for example, in an RHC with FM-AFM coupling.\cite{Westerberg95} 
For this reason, the onset of saturation of the low-temperature magnetization 
occurs at $k_{\mathrm{B}} T_i \approx \mu_{\mathrm{B}} H$.\cite{Westerberg95} 
A fit to the susceptibility data at the smallest applied field, using the theoretical 
random-singlet expression for $\chi(T)_{\mathrm{RHC}}$, is shown in 
Fig.~\ref{fig:RHC_chi_meas} as a black dashed line. A rather good match is found 
above 4 K with a cutoff $J_0 \simeq 66.3 \pm 0.7$ meV.

To achieve a more quantitative comparison, we carried out a series of Quantum Monte Carlo 
(QMC) simulations based on Eq.~\eqref{rhc1}, where $J_i$ couplings alternate randomly from 
$J_1 = J_{\mathrm{Si}} = 24.1$~meV to $J_2 = J_{\mathrm{Ge}} = 50$~meV 
(we call this a $J_1$--$J_2$ model), mimicking the situation 
expected for \BASGE.
We chose 6000 spin sites with randomly distributed but equally probable $J_1$ and $J_2$ 
couplings. The simulations for $\chi(T,H)$ were averaged over at least 25 random realizations 
of disorder, as obtained from a directed-loop algorithm \cite{Alet2005} within the ALPS 2.0 
package.\cite{Bauer2007} 
Technical details related to the QMC simulations are discussed in App.~\ref{app:A}.
Disorder in the exchange couplings was imposed by assuming an equal number 
($L/2$, with $L$ as the chain length) of $J_{1}$ and $J_{2}$ values, randomly permuted 
to construct the RHC chain that was to be simulated. 

Selected results of these simulations are shown as red and yellow circles in 
Fig.~\ref{fig:RHC_chi_meas}, as interconnected diamonds in Fig.~\ref{fig:Phase_bound_cuts}, 
and as a color map in Fig.~\ref{fig:Phase_bound}b. In view of the simplicity of the model, 
the qualitative and quantitative agreement of our parameter-free QMC simulations with 
the data is remarkable. 
Incidentally, the best fit of the RS prediction to the low-$T$ QMC results in an applied 
field $\mu_0 H = 0.1$~T requires $J_0 \approx J_2$. This corresponds to the largest energy 
scale in the system before the decimation starts, thus reinforcing the pertinence of the 
theory used to capture the magnetic properties of the system. 
As may be seen in Figs.~\ref{fig:Phase_bound_cuts} and \ref{fig:Phase_bound}b, by applying 
the moving-average procedure described above also to the QMC results, numerical calculations 
predict the same free-spin-like linear field-dependence for $T_i$ as observed experimentally. 
This is an important result, because the existence of paramagnetic entities at $T = 10$ K 
in a strongly-correlated system, with a characteristic energy scale corresponding to a 
temperature exceeding 200 K, is highly counterintuitive.

\section{Nuclear Magnetic Resonance in a Random Heisenberg Chain}
\label{sec:nmr1}
Although the measurements of $\chi(T,H)$ and the comparison with QMC calculations already provide 
a strong indication for the formation of an RS phase in \BASGE\ at low temperatures, it is tempting to 
directly verify the presence of a random-singlet state on a local scale.
With decreasing temperature an increasing number of spins form pairs by adopting a 
nonmagnetic singlet ground state. The divergent low-temperature tail of $\chi(T,H)$ 
is due to the residual unpaired spins. The tendency to form a random-singlet ground 
state should be directly manifest in NMR data, reflecting a growing number of $^{29}$Si 
sites which experience a zero net transferred local field. 

\subsection{Experimental results}
\begin{figure}[th!]
\centering
\includegraphics[width=0.45\textwidth]{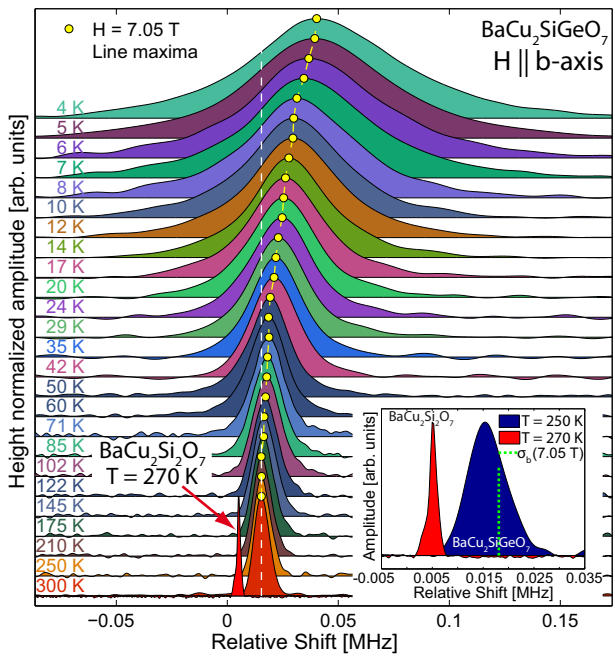}  
\caption{\label{fig:NMR_Lines}(color online) Temperature dependence of the $^{29}$Si NMR 
lines of the disordered Heisenberg chain compound \BASGE\ in a magnetic field of 7.05 T 
applied along the $b$-axis. The zero frequency marks the resonance of $^{29}$Si nuclear 
spins in a standard reference sample such as Si(CH$_3$)$_4$.\cite{HARRIS01} The white 
dashed line marks the position of the resonance at room temperature, while the yellow 
points represent the position of the maxima, displaying a shift towards higher frequencies 
at low temperatures. The inset compares the cases of maximum and no disorder at high 
temperature. The green dashed line marks the orbital shift as obtained from Ref.~\onlinecite{Casola12} 
(see text for details).}
\end{figure}

Height-normalized $^{29}$Si NMR lines, as recorded in a 7-T external field applied along 
the crystallographic $b$ axis, are plotted in Fig.~\ref{fig:NMR_Lines}. The spectra were 
obtained by superposing several acquisitions in the frequency-sweep mode, as described 
in Ref.~\onlinecite{Clark95}.

The effects of disorder are evident already by comparing the high-temperature line shapes 
of \BASGE\ and \BACSO\ (see inset of Fig.~\ref{fig:NMR_Lines}). We note that the $^{29}$Si 
resonance of \BASGE, apart from being broader than that of its pristine counterpart, is also 
shifted to higher frequencies. We recall that NMR data of \BACSO\ (taken at $\mu_0 H = 7$ T, 
with $\mathbf{H} \parallel b$) exhibit a positive orbital shift of $\sigma_b \simeq 0.018$ MHz, 
with a \textit{negative} hyperfine-coupling to the longitudinal magnetization.\cite{Casola12}
As shown in Fig.~\ref{fig:RHC_chi_meas}, the high-temperature magnetization of \BASGE\ is 
equivalent to that of an isotropic chain with $J=37$ meV and is distinctly smaller than that of 
\BACSO, well modelled by choosing $J=24.1$ meV. Consequently, the negative hyperfine shift 
in case of a sample with disorder is smaller and, hence, the resonance is located closer to 
$\sigma_b$ (see inset of Fig.~\ref{fig:NMR_Lines}). This result is ultimately an independent 
confirmation of the validity of the model which allowed us to extract the parameter $\sigma_b$ 
in Ref.~\onlinecite{Casola12}.
We emphasize that, due to a random variation of the hyperfine interactions, a broadening 
of the resonance in the $x=0.5$ case  may well be of structural origin, rather than due to 
magnetic disorder. Realistically, both structural and magnetic disorder have to be taken 
into account and one of the aims of the analysis outlined below is to disentangle them. 
Another striking feature of the NMR signals is the temperature-dependent shift of the 
line maxima. In Fig.~\ref{fig:NMR_Lines} this shift is evidenced by comparing the sequence 
of yellow points, marking the maxima, with the vertical dashed line, representing the peak 
position of the 300-K line. It indicates a growing nonzero uniform magnetization as the 
temperature is lowered, in obvious contradiction to the RS-phase hypothesis, where the 
formation of singlets would imply the absence of a local magnetization. 
As expected, unlike to what is observed in \BACSO,\cite{Kenz01,Casola12} the NMR data 
show no indications of a phase transition around $T = 10$ K for the disordered compound.
\begin{figure*}[th]
\centering
\includegraphics[width=0.7\textwidth]{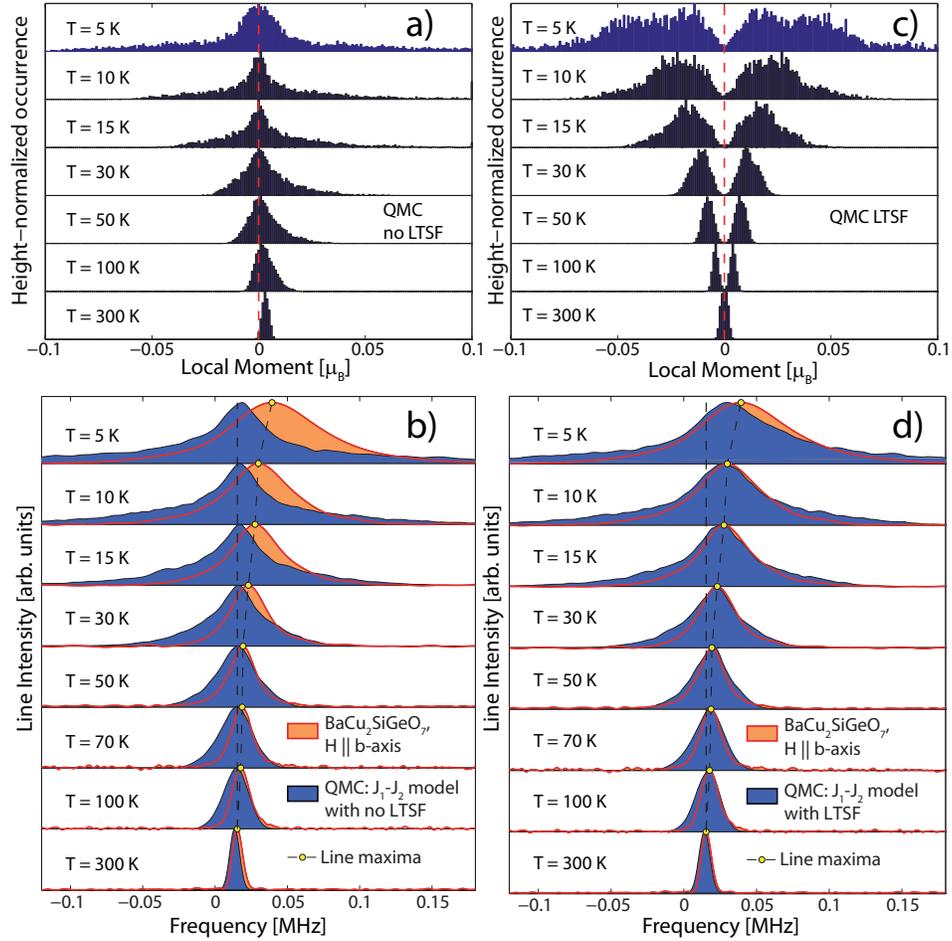}   
\caption{\label{fig:Lines_NMR_simulation}(color online) (a) QMC simulation of the local 
magnetization of a $J_1$--$J_2$ model of a spin-chain with 6000 sites placed in an external 
field of 7 T. The number of occurrences of the magnitude of a certain local moment is plotted, 
with each occurrence calculated in equidistant intervals of $10^{-3} \mu_{\mathrm{B}}$ and 
normalized to the most frequent one. The red dashed line marks the zero. (b) Simulated NMR 
line shapes (blue) using the results obtained from (a) and compared with the experimental 
data (orange). The black dashed line at fixed frequency marks the position of the peak of 
the \BASGE\ line at 300 K. No LTSF is present in this case. (c) Same as in (a), but with the 
uniform field replaced by a staggered field $H_{i,\perp} = (-1)^icH$, with $c = 0.092$ and 
$\mu_0 H = 7$ T. (d) Same as in (b), but including an LTSF in the model (see text for details).}
\end{figure*}

\subsection{\label{ssec:nmr_analysis}Data analysis}
In order to interpret the recorded NMR line shapes and positions, we compare the experimental 
data with the results of QMC simulations of the $J_1$--$J_2$ model, addressed in Sec.~\ref{sec:magnQMC}. 
In Fig.~\ref{fig:Lines_NMR_simulation}a we show the temperature-dependent, height-normalized 
histograms, which reflect the occurrences of longitudinal local magnetization values $S^z_i$ in 
$10^{-3}$ $\mu_{\mathrm{B}}$ intervals for $\mu_0 H = 7$ T ($z$ is the spin quantization axis 
along which the external field was applied). The histograms were obtained by using a single 
configuration of disorder in a system of $L = 6000$ spins. 
We note that at $T = 300$ K the $S^z_i$ values are distributed around a nonzero, albeit small, uniform 
magnetization. In this regime the details of disorder are not essential for describing the physics. At 
low temperatures, however, a broadened magnetization peak appears, centered at zero local 
moment, as expected in the case of a continuous formation of singlets. At the same time we observe 
weak but extremely broad tails (see top histogram in Fig.~\ref{fig:Lines_NMR_simulation}a), indicating 
the presence of incompletely compensated spins, which are progressively polarized as the 
temperature is reduced.

As shown in Fig.~\ref{fig:Lines_NMR_simulation}b, the local-moment distribution 
obtained from these simulations (which do not include an LTSF) does not match 
the experimental NMR data. 
The formation of random singlets implies that at a significant number of sites 
the moments of the Cu atoms are quenched at low temperatures, corresponding 
to a zero shift of the simulated NMR line, in disagreement with observations. 
It turns out that it is essential to consider the combined effect of random exchange 
interactions and an LTSF, namely to set $c \neq 0$ in the Hamiltonian \eqref{rhc1}.  
To include an LTSF we use the same approximation made in the case of no disorder,\cite{Casola12,Affleck99} 
i.e., we assume the total local magnetization $\mathbf{m}_{i}$ to be the sum of a non-uniform 
longitudinal and a transverse contribution, in the form:
\begin{equation}
\label{Feiguin_renormalized}
\mathbf{m}_{i} = \langle \mathbf{S}^{\parallel}_i \rangle + \langle \mathbf{S}_i^{\perp} \rangle,
\end{equation}
where for the magnitude of the local transverse magnetization we write 
$\langle S_i^{\perp} \rangle(H, H_{i,\perp}) \approx \langle S_i^{\perp} \rangle(0, H_{i,\perp})$, and analogously for the longitudinal component, with $H_{i,\perp} = c_i H$.\cite{Casola12}
\begin{figure}[th]
\centering
\includegraphics[width=0.48\textwidth]{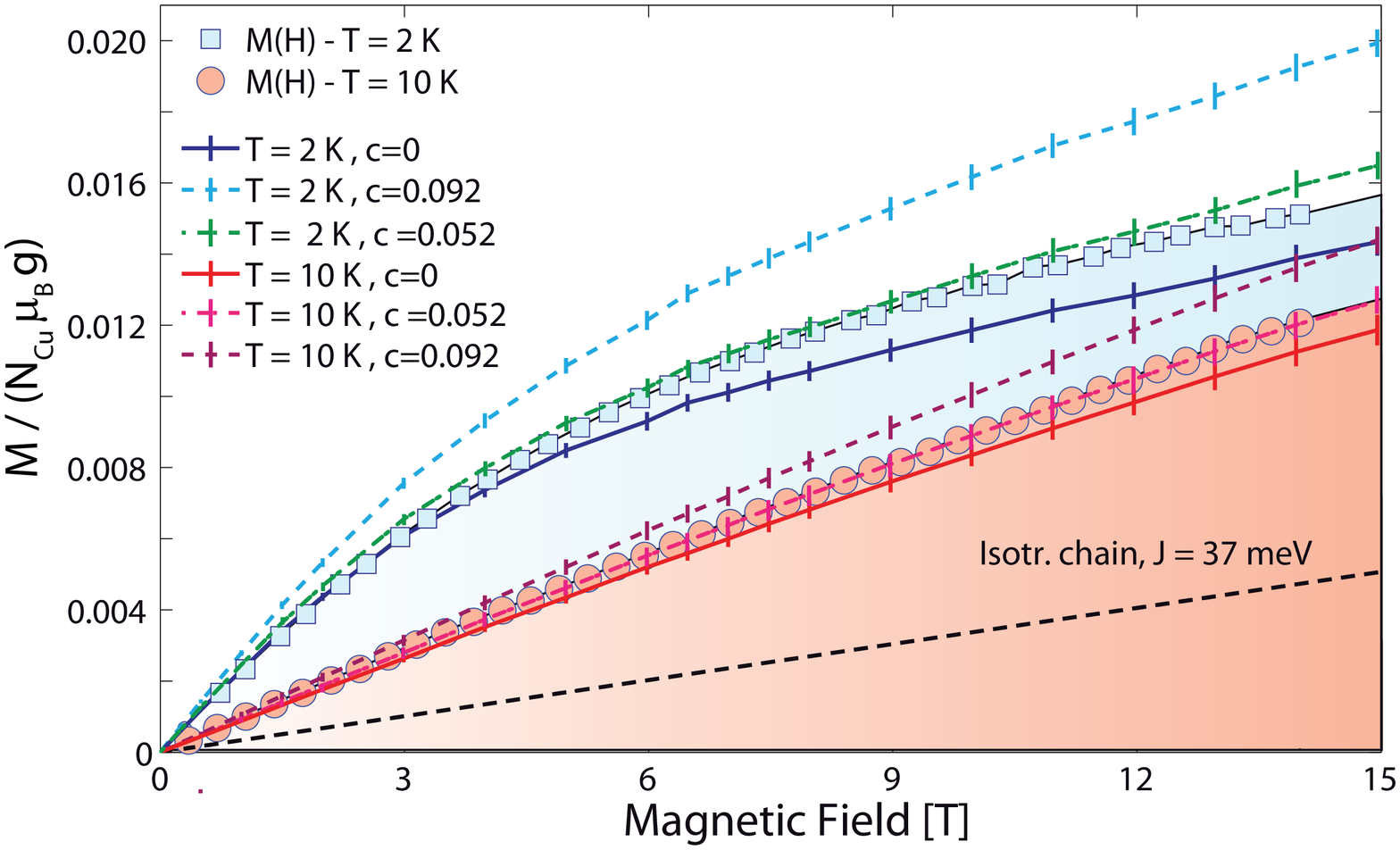}  
\caption{\label{fig:M_H_qmc}(color online) Field dependence of the magnetization of \BASGE\ at 
2 and 10 K (dots) and comparison with QMC simulations of the $J_1-J_2$ RHC model without 
($c=0$) and with ($c\neq0$) LTSF. The $c$ coefficient refers to the staggered field value 
$H_{i,\perp} = (-1)^icH$, with $H$ the uniform field. Also shown is the field dependence of 
the ground state magnetization ($T=0$ K) of an isotropic Heisenberg spin-\nicefrac{1}{2} 
chain, as from Ref.~\onlinecite{Johnston00}. The magnetization is expressed in normalized 
units ($N$ is the total number of copper sites and $\mu_{\mathrm{B}}$ is the Bohr magneton), 
assuming $g=2$.}
\end{figure}
Because of this particular approximation, two QMC simulations, one for $S^{\perp}_i$ 
and another for $S_i^{\parallel}$, can be carried out separately on a diagonal basis.\\
With $H \neq 0$, an LTSF acts on paired spins by mixing the nonmagnetic singlet states $|s\rangle$ 
with the magnetic triplet states $|t_{+1}\rangle = |\uparrow_1\uparrow_2 \rangle$ and 
$|t_{-1}\rangle = |\downarrow_1\downarrow_2 \rangle$, thereby inducing a nonzero magnetization 
also in the ground state of the spin pair.\cite{Miyahara07} If both the uniform field and the LTSF 
are considered, in a first-order perturbation approximation, the new ground state can be written 
as:\cite{Miyahara07} 
\begin{align*}
\label{eq:perturb}
|s'\rangle & \approx |s\rangle + \alpha_+ |t_{+1}\rangle - \alpha_- |t_{-1}\rangle, \quad \mathrm{where}\\
\alpha_{\pm} & \propto \frac{cH\mu_{\mathrm{B}}} {J \mp g \mu_{\mathrm{B}} H}.
\end{align*}
The uniform field appearing in the denominator can be neglected if $J \gg g \mu_{\mathrm{B}} H$, 
which is the case for the strongest coupled singlets in the RS phase. Therefore, at temperatures 
$k_{\mathrm{B}} T \approx \min(J) \gg g \mu_{\mathrm{B}} H$ the longitudinal local magnetization 
varies linearly with the applied field. 
Here min($J$) refers to the minimum exchange coupling value of the already formed singlets.

The distribution of local-magnetization values in a $J_1$--$J_2$ RHC model which includes an LTSF 
is shown in Fig.~\ref{fig:Lines_NMR_simulation}c. Clearly, the presence of an LTSF substantially 
modifies the magnetization profiles shown in Fig.~\ref{fig:Lines_NMR_simulation}a. By assuming 
that the local symmetry of the oxygen atoms surrounding a Cu$^{2+}$ ion is largely unaffected 
by the Si-to-Ge substitution, the same value of $c=0.092$ obtained for \BACSO\ ought to be 
valid also in our case.
As seen in Fig.~\ref{fig:Lines_NMR_simulation}c, at $T = 300$ K and $\mu_0 H = 7$ T, the chosen 
LTSF is too weak to induce a nonzero magnetization. However, as the temperature is progressively 
lowered, singlets with a \textit{distribution} of effective exchange couplings are formed and a 
staggered field $H_{i,\perp}$ acting on them creates a distributed local magnetization of the 
form $\langle S^{\perp}_i \rangle \propto \mu_{\mathrm{B}} H_{i,\perp}/J_i$.
Therefore, the histograms in Fig.~\ref{fig:Lines_NMR_simulation}c split and broaden with 
decreasing temperature. The symmetrical splitting reflects the alternating sign of the LTSF 
from site to site, with a mean absolute value of the local moment which progressively shifts  
away from $|\langle S^{\perp}_i \rangle|=0$.

In order to compare the simulated local magnetization profiles with the NMR line shapes, 
structural details have to be considered. Eight configurations of disorder in systems with 
$L = 6000$ sites were simulated. In order to reproduce 6000 unit cells, the local moments 
were then arranged into four different chains, with two inequivalent copper atoms each.
At this first stage, the dipolar coupling was neglected, but we considered the hyperfine 
coupling 
to both the longitudinal and the transverse magnetization. Following the notation of 
Ref.~\onlinecite{Casola12}, for $\mathbf{H} \parallel b$, the local hyperfine field 
$h_{\mathrm{loc}}$ at a general $^{29}$Si nuclear site can be written as:
\begin{equation}
h_{\mathrm{loc}} = \sum_{i=1}^4 \Big( a_i \langle S^{\parallel}_i \rangle + b_i \langle S^{\perp}_i \rangle \Big),
\end{equation}
where the index $i$ runs over the four nearest-neighbor copper sites. With this notation, 
the relative NMR resonance frequency $^{29}\omega$ can be written as 
$^{29}\omega \approx \gamma h_{\mathrm{loc}} + \sigma_b$, with $\gamma$ as the $^{29}$Si 
nuclear gyromagnetic ratio. From our previous study\cite{Casola12} of \BACSO\ we can fix the 
parameters $\sum_i a_i \simeq -0.16$ T/$\mu_{\mathrm{B}}$, $\sum_i (-1)^{i} b_i \simeq 0.128$ T/$\mu_{\mathrm{B}}$, 
whereas $\sigma_b \simeq 0.018$ MHz at 7 T.
Each of the above hyperfine parameters, which couple the silicon nuclear magnetism to 
the longitudinal and transverse electronic magnetization of copper, represents the sum 
of four individual copper-to-silicon couplings. Given the broken translational symmetry due 
to disorder, the individual Cu-Si couplings are essential for a quantitative comparison 
between data and theory. Unfortunately, these cannot be evaluated by experiment, leaving 
six unknown free parameters in the model. 

For a comparison of the simulated NMR lines with data recorded at 5 K, we used 
the following parameters (in T/$\mu_{\mathrm{B}}$ units): $b_1 = b_3 = -0.24$, 
$b_2 = -0.07$ for the coupling to the transverse magnetization, and $a_1 = -0.16$, 
$a_2 = 0.08$, $a_3 = -0.015$ for the coupling to the longitudinal one. 
The sign of the parameters $b_i$ changes according to the considered Si nucleus.
By making use of them and with no other free parameters, we succeeded in simulating the 
complete temperature dependence of the NMR lines, which account for the simultaneous 
presence of both an LTSF and a uniform magnetic field. The result is shown in 
Fig.~\ref{fig:Lines_NMR_simulation}d, where the agreement between data and the 
theoretical model is again remarkable. 

To be more specific, for a comparison with the experiment, we calculated the line shapes
by convoluting the distribution of local fields at the $^{29}$Si sites with a 2-kHz wide Gaussian, 
which is equivalent to the width of the line profile of pure \BACSO, shown in the inset of 
Fig.~\ref{fig:NMR_Lines}. The choice of the optimal binning range, which strongly affects 
the shape of the tails of the $S^z_i$ histograms in Fig.~\ref{fig:Lines_NMR_simulation}, 
was done by employing the algorithms proposed in Refs.~\onlinecite{Shimazaki07,Shimazaki07V2}. 
Note that the lines displayed in Fig.~\ref{fig:Lines_NMR_simulation}d show that the averaging 
process over four Cu sites removes the splitting in the original histograms in Fig.~\ref{fig:Lines_NMR_simulation}c. 
In addition, the simulated lines shift with temperature, as indeed observed in the experiment. 
The minimally-shifted line shape at $T = 300$ K reflects magnetic disorder. 
By comparing Figs.~\ref{fig:Lines_NMR_simulation}b and \ref{fig:Lines_NMR_simulation}d, 
it may be seen that the inclusion of an LTSF in the simulations significantly improves the 
agreement between theory and experiment.
Therefore, from both a macroscopic and a local point of view, NMR data confirm that the static 
magnetic properties of \BASGE\ are those of an RHC, with the addition of a residual LTSF.

To cross check the consistency of the proposed model, it seems natural to consider 
the LTSF-induced effects also in the magnetometry data. Since the influence of the LTSF
 grows with the applied field $H$, we consider the calculated $M(H)$ curves (shown by 
dotted lines in Fig.~\ref{fig:M_H_qmc}) at two representative temperatures, $T = 2$ K 
and 10 K.
The calculation of  $M(H)$ at $T = 2$ K, using the average exchange coupling $J = 37$ meV, 
reflects the \textit{linear} behavior expected for an isotropic Heisenberg chain in its LL 
regime\cite{Johnston00} and differs substantially from the experimental data.
In fact, this oversimplified model completely neglects the field-induced alignment of 
the still uncompensated spins that, in a random Heisenberg chain, survive even at 
the lowest temperatures. 
On the other hand, QMC simulations using the $J_1$--$J_2$ model discussed in 
Sec.~\ref{sec:magnQMC} and in Ref.~\onlinecite{Shiroka11} show significant improvements 
over the simplified ``average-$J$'' model (solid blue and red lines in Fig.~\ref{fig:M_H_qmc} 
for $T = 2$ K and 10 K, respectively).

The next refinement was to include both a uniform and a staggered magnetization
in the $J_1$--$J_2$ model and combine them\cite{Affleck99} to obtain:
\begin{equation}
 M(H,T) = \sum_i \Big[ \langle S^{\parallel}_i \rangle(H,T) + (-1)^i c  \langle S_i^{\perp} \rangle(H_{i,\perp},T) \Big], 
\end{equation} 
with $\sum_i \langle S^{\parallel}_i \rangle(H,T)$ the RHC magnetization in a uniform field. 
As can be seen in Fig.~\ref{fig:M_H_qmc}, the QMC calculation for $M(H,T)$ with $c = 0.092$ 
unexpectedly fails to reproduce the data. A good agreement with the magnetization scans 
at both temperatures was, however, obtained using $c \simeq 0.052$. 
The discrepancy in the prefactor is most probably due to the fact that the current model 
neglects the effects of structural disorder in \BASGE. A more refined model taking into 
account such subtleties, by including a site-dependent $c$ value, was not implemented in 
the present analysis. 
Nonetheless, the essence of RHC physics, which dominates the magnetic properties of 
the probed system, is born out by the combined investigation of bulk magnetometry, 
NMR measurements and QMC simulations. 
Taken together they strongly suggest that \BASGE\  is an excellent realization of 
a random Heisenberg chain model, as originally claimed in Ref.~\onlinecite{Shiroka11}, 
thus providing the ideal testing ground for analytic theories modeling a random-singlet 
ground state. 

\section{Summary and Conclusions}
\label{sec:summ}
In conclusion, we considered the \BASGE\ compound, where spin-$\nicefrac{1}{2}$ Cu$^{2+}$ 
ions interact via random exchange couplings, as a representative of RHC systems with bond 
disorder. We demonstrate that this spin-chain material has a ground state which is very well 
described on the basis of the RHC Hamiltonian in Eq.~\eqref{rhc1}.

By using temperature- and magnetic field-dependent microscopic ${}^{29}$Si NMR measurements, 
combined with macroscopic magnetic-susceptibility data and results of detailed numerical Quantum 
Monte Carlo simulations, we find a coherent description of the physical properties of the system, 
compatible with that of an RS phase.
In particular, the considerable broadening of the NMR lines and the divergent magnetic response 
at low temperatures, combined with a nonlinear increase of the magnetization as a function of 
field at low $T$, all reflect the presence of unpaired spins with arbitrary small couplings even 
at temperatures close to zero.

Specific refinements in the simulations of $J_1$-$J_2$ model have shown that the addition 
of a local transverse staggered field --- LTSF (similar to that present in the parent compound 
\BACSO)\cite{Casola12} --- is essential also for reproducing  the observed NMR line shifts.

Since the magnetization of a singlet in an LTSF is proportional to the local field, the detection 
of the field distribution by future ${}^{63}$Cu NMR measurements may provide a direct way 
to explore the $J$ distribution in an RHC. In addition, besides the existing stretched-exponential 
NMR relaxation data,\cite{Shiroka11} the spectral properties of random Heisenberg chains 
would be more directly accessible if neutron scattering experiments in a lower energy range 
would be possible.

\begin{acknowledgments}
The authors thank A.\ Feiguin (Northwestern University, Boston), K.\ Pr\v{s}a (EPFL, Lausanne), and 
O.\ Zaharko (PSI, Villigen) for useful discussions. The \BASGE\ samples used in this work were 
prepared in the early 2000's in the group of Prof.\ K.\ Uchinokura at the University of Tokyo. 
This work was financially supported in part by the Schweizerische Nationalfonds zur F\"{o}rderung 
der Wissenschaftlichen Forschung (SNF) and the NCCR research pool MaNEP of SNF.
\end{acknowledgments}

\appendix
\section{\label{app:A}Error evaluation in the QMC simulations}
In order to correctly interpret the numerical results, it is necessary to mention the theoretical 
expectations for the error bars in QMC simulations (i.e., specific to this method), in relation 
to the self-averaging induced errors (i.e., specific to disorder).\cite{Aharony96} 
For a spin chain of length $L$, one can define the ratio 
$R_X(L) = (\overline{X^2} - \overline{X}^2)/ \overline{X}^2$, where $X$
is any macroscopic variable subject to self-averaging and $\overline{X}$ denotes 
the same quantity averaged over the realizations of disorder.\cite{Aharony96} 
By using $10^4$ ``Monte Carlo updates" (the number of random updates of the system performed 
by the algorithm before a new configuration is generated) and $10^5$ independent Monte Carlo 
measurements in a system of 6000 spins, the relative error for $M(T,H)$ in low-temperature 
QMC simulations was found to be 10 times lower than $\sqrt{R_X(L)}$, the error due to the average 
on all the realizations of disorder.
Therefore,  in our case $\sqrt{R_X(L)}$ is the dominant simulation error. Whenever the correlation 
length is $\xi \sim \log^2(J_0/ T) \ll L$, self-averaging can be justified based on the central-limit 
theorem.\cite{Aharony96}
In case of critical systems, where the correlation length tends to diverge ($\xi \gg L$), the only 
way to obtain meaningful results in the presence of disorder consists in increasing the system 
size $L$. By simulating systems with different spin-chain lengths $L$, we find that 
$\sqrt{R_X(L \rightarrow \infty)} = \sigma_{\chi}/(\overline{X}\sqrt{L})$, with $\sigma_{\chi}$ the 
standard deviation of the local spin susceptibility. This implies the correct evaluation of the 
thermodynamic limit even in the problematic $\xi \gg 1$ case. Incidentally, since $\sigma_{\chi}$ 
is smaller at higher temperatures and magnetic fields, this justifies the smaller error bars 
in these regions.

\end{document}